\documentclass[twocolumn]{aastex63}
\usepackage{amsmath}
\usepackage{color}
\def\beq{\begin{eqnarray}}
\def\eeq{\end{eqnarray}}

\begin{document}

\title{Stochastic gravitational wave background from cosmological neutrino-dominated accretion flows}

\correspondingauthor{Tong Liu}
\email{tongliu@xmu.edu.cn}

\author[0000-0002-9130-2586]{Yun-Feng Wei}
\affiliation{Department of Astronomy, Xiamen University, Xiamen, Fujian 361005, China}

\author[0000-0001-8678-6291]{Tong Liu}
\affiliation{Department of Astronomy, Xiamen University, Xiamen, Fujian 361005, China}

\begin{abstract}
We investigate the stochastic gravitational wave background (SGWB) from neutrino-dominated accretion flows (NDAFs) based on the results of our fallback core-collapse supernova (CCSN) simulations. We find that the predicted SGWB is mainly determined by the typical CCSN initial explosion energy and progenitor metallicity. For the optimistic cases in which the typical initial explosion energy is low, the SGWB from NDAFs without disk outflows might be detected by next-generation space-based interferometers such as Decihertz Interferometer Gravitational wave Observatory (DECIGO) and Big Bang Observer (BBO). In the low-frequency regime $\sim10^{-3}-10^{-1}$ Hz, this background is comparable to that expected from standard inflationary models. Therefore, the SGWB from NDAFs may become a foreground for searches of the SGWB generated in the inflationary epoch. Combining the diffuse NDAF neutrino background and SGWB from NDAFs, one may constrain the properties of the CCSNe and NDAFs.
\end{abstract}


\section{Introduction}

The stochastic gravitational wave background (SGWB) is expected to be created by the superposition of gravitational waves (GWs) from many uncorrelated sources. Such a background could be produced in many processes during cosmological and astrophysical evolutions. A cosmological SGWB may be produced by inflation \citep[e.g.,][]{Grishchuk1975,Starobinski1979,Easther2006,Smith2006,Kuroyanagi2009,Barnaby2012}, cosmic strings \citep[e.g.,][]{Caldwell1992,Damour2000,Damour2005,Siemens2007,Olmez2010,Olmez2012}, alternative cosmologies \citep[e.g.,][]{Gasperini1993,Mandic2006}, and a variety of other phenomena. Astrophysical SGWB could have various possible sources (see \citet{Regimbau2011} for a review ), including inspiral and coalescence of compact binaries \citep[e.g.,][]{Phinney1991,Kosenko1998,Regimbau2006,Zhu2011a,Rosado2011,Marassi2011,Wu2012,Zhu2013,Evangelista2015,Kowalska2015,Evangelista2015}, white dwarf binaries \citep[e.g.,][]{Farmer2003}, rotating neutron stars \citep[NSs, e.g.,][]{Regimbau2001,Ferrari1999,Houser1994,Lai1995,Zhu2011b,Lasky2013}, core-collapse supernovae \citep[CCSNe, e.g.,][]{Ferrari1999}, gamma-ray bursts \citep[GRBs, e.g.,][]{Hiramatsu2005}, Population III (Pop III) stars \citep[e.g.,][]{Suwa2007,Marassi2009}, stellar core collapse \citep[e.g.,][]{Crocker2015,Crocker2017}, and so on.

The next-generation space GW detectors, such as Decihertz Interferometer Gravitational wave Observatory \citep[DECIGO,][]{Seto2001} and Big Bang Observer \citep[BBO,][]{Ungarelli2005}, targeting $0.1-1$ Hz, might be possible to detect SGWB from both cosmological and astrophysical origins. Notably, these experiments are expected to detect the primordial SGWB that arises in the very early universe during the inflationary epoch \citep{Maggiore2000}. Such SGWB may serve as a powerful tool for studying the extremely early universe. However, astrophysical foreground sources could be a significant problem for searches of the inflationary SGWB. The SGWB from CCSNe \citep{Buonanno2005} and Population III stars \citep{Suwa2007,Marassi2009} are expected to be comparable to or mask the inflationary SGWB in some range of frequencies. Thus, a good understanding of astrophysical foregrounds around the deci-Hertz band is essential for DECIGO and BBO detecting the inflationary SGWB. In this paper, we focus on the low-frequency GW generated by anisotropic neutrino emission of neutrino-dominated accretion flows (NDAFs) and estimate the SGWB spectrum from NDAFs.

NDAFs around rotating stellar-mass black holes (BHs) are one of the plausible candidates of GRB central engines in massive collapsars and compact object mergers. In the collapsar model \citep[e.g.,][]{Woosley1993,MacFadyen1999} for long-duration GRBs, the core collapse of a massive star can produce a BH hyperaccretion system via the fallback process. For the very high accretion rate, the disk might be in a state of NDAFs. In the inner region of the disk, photons are trapped, and only neutrinos are emitted from the disk surface. These neutrinos annihilate in the space out of the disk and then form the primordial fireball to power a GRB. The properties of NDAFs have been widely studied in recent decades \citep[see e.g.,][]{Popham1999,Narayan2001,Kohri2002,Lee2005,Gu2006,Chen2007,Janiuk2007,Kawanaka2007,Liu2007,Liu2014,Lei2009,Xue2013,Song2016}, and for a review see \citet{Liu2017a}. In particular, the GW signals generated by the anisotropic neutrino emission from NDAFs have been investigated by some previous works \citep[e.g.,][]{Liu2017b,Wei2020,Song2020,Wei2021,Chen2022,Qi2022}. The neutrino-induced GWs are detectable for $\sim10$ Mpc by DECIGO/BBO \citep{Suwa2009}. Of course, the BZ mechanism \citep{Blandford1977} could coexist even be dominated in the BH hyperaccretion process, but it has no GW radiation.

An earlier estimation of the SGWB produced by NDAFs was given by \citet{Suwa2009}. They adopted some typical GW energy spectra of NDAFs to calculate the SGWB. Meanwhile, they used the GRB formation history to estimate the event rate of NDAFs. The results show that the SGWB from NDAFs is below the detection limit of DECIGO/BBO. However, the effects of progenitor properties and the initial explosion energy on the GW spectrum of NDAF are not included in their work. Besides, the event rate of NDAFs might be underestimated. In the collapsar model, an observable GRB is only triggered when jets launched by the central engine can break out from the envelope and circumstellar medium in the prompt emission phase \citep{Wei2022}. If the jet is chocked in the envelope or the observer is not in the jet's line of sight, no GRB would be detected. Therefore, the event rate of GRBs should be much less than that of NDAFs. Of course, the GWs from NDAFs are partly anisotropic. In this work, we estimate the SGWB from NDAFs based on fallback CCSN simulations. In \citet{Wei2021}, we revealed how the GW emission of NDAFs depends on the initial explosion energies, masses, and metallicities of the progenitor stars. Here, we aim to improve the estimation of the SGWB from NDAFs by including those dependencies.

In this work, we mainly focus on the SGWB from NDAFs around BHs. Note that NDAFs around NSs may also operate in some GRBs and CCSNe \citep[e.g.,][]{Zhang2010,Perna2014}. Actually, for the newborn NSs in the center of massive collapsars, their strong magnetic fields \citep[$\gtrsim 10^{15}~\rm G$, see e.g.,][]{Song2023} will restrain the accretion process and destroy the inner region of the disk. Besides, since the rapidly decreasing fallback accretion rate, NDAFs might only exist for the very initial stage of the NS hyperaccretion process. Therefore, the contribution of NDAFs around NSs to the SGWB is not considered here.

The paper is organized as follows. In Section 2, we describe the setup of our CCSN simulations and discuss the effects of the initial explosion energies and the masses and metallicities of the progenitor stars on the GW emission of NDAFs. In Section 3, we estimate the amplitude of the SGWB from NDAFs. We discuss how the SGWB depends on the initial explosion energy and progenitor metallicity. We also explore the effect of initial mass function (IMF) on the SGWB from NDAFs. The conclusions and discussion are presented in Section 4.

\section{Model}

\subsection{CCSN simulations}

Here, we briefly review the setup of our CCSN simulations. In this paper, we adopt the presupernova (pre-SN) models with initial mass in the range of $20-40$ $M_{\odot}$ as progenitor models \citep[e.g.,][]{Woosley2002,Woosley2007,Heger2010}. For those models, the models with zero metallicity ($Z/Z_{\odot} = 0$) and solar metallicity ($Z/Z_{\odot} = 1$) are referenced from \citet{Heger2010} and \citet{Woosley2007}, respectively, as well as the ones with metallicity $Z/Z_{\odot} =0.01$ are provided by Prof. Alexander Heger in private communication, where $Z_\odot$ represents the metallicity of the Sun. The stellar collapse and explosion simulations \citep{Liu2021,Wei2022} are performed in a series of 1D simulations with the Athena++ code \citep{White2016}. We use the piston approach \citep{Woosley1995,Woosley2002} to carry out spherically symmetric explosion simulations. For each progenitor star, a piston was initially located at the outer edge of the iron core. When the star collapses, the piston firstly moves inward for 0.45 s and then moves outward with an initial high velocity and decelerates smoothly until coming to rest at $10^{9}$ cm. We follow this approach to determine the initial explosion condition at the inner boundary.

The simulation was divided into two steps to reflect the initial collapse and the subsequent explosion. In the first step, the numerical grid has an inner boundary at $10^{9}$ cm, and a unidirectional outflowing inner boundary condition was used at the inner boundary to mimic the suction effect resulting from the hypothetical piston moving inwards. The outer boundary is set at the surface of the progenitor star. In this step, the grid has $10^{4}$ cells with a logarithmic uniform interval for the radial direction. The simulation is run to 0.45 s, and then the piston turns outwards, corresponding to the outward propagation of the blast.

In the second step, we use the same outflowing inner boundary condition, which is set at $10^{9}$ cm. The outer boundary is set at $10^{16}$ cm, and the medium outside the star is maintained in a constant state with a pressure and density three orders of magnitude lower than the corresponding ones on the star surface. In this step, the grid has 2,000 logarithmic cells. At the beginning of this step, we map the results of the first step to the new grid for the second step and inject energy into the innermost cell adjacent to the inner boundary to mimic the outward blast passing through the inner boundary. Here, the injected energy is the setting energy, which is assumed to have three values for each case, i.e., 2, 4, and 8 $\rm B$ ($1{\rm B}=10^{51}~\rm erg$). All simulations were run until the remnant growth ceased. For more details of simulations, see \citet{Liu2021} and \citet{Wei2022}.

For each simulation, we record the evolution of the fallback mass supply rate at the inner boundary. Ignoring the disk outflows, we roughly consider the mass supply rate as the mass accretion rate of the disk. If the mass accretion is high, the hyperaccretion disk would be in the state of NDAFs. Meanwhile, the mass and spin of the BH will violently evolve within tens of seconds. According to the conversion of the energy and angular momentum, the evolution equations of a spinning BH can be expressed as \citep[e.g.,][]{Hou2014,Song2015},
\beq
\frac{dM_{\rm{BH}}}{dt}=\frac{\dot{M}}{\sqrt{3x_{\rm{ms}}}}\left ( 4-\frac{3a_{*}}{\sqrt{x_{\rm{ms}}}} \right ),
\eeq
and
\beq
\frac{da_{*}}{dt}=\frac{2\sqrt{3}\dot{M}}{M_{\rm{BH}}}\left ( 1- \frac{a_{*}}{\sqrt{x_{\rm{ms}}}}\right )^{2},
\eeq
where $M_{\rm{BH}}$ and $\dot{M}$ are the mass of the BH and the mass accretion rate, respectively, $a_{*}$ is the dimensionless spin parameter of the BH, and $x_{\rm{ms}}=3+Z_{2}-\sqrt{(3-Z_{1})(3+Z_{1}+2Z_{2})}$ is the dimensionless radius of the marginally stable orbit \citep{Bardeen1972,Kato2008}, where $Z_{1}=1+(1-a_{*}^{2})^{1/3}[(1+a_{*})^{1/3}+(1-a_{*})^{1/3}]$ and $Z_{2}=\sqrt{3a_{*}^{2}+Z_{1}^{2}}$ for $0< a_{*}< 1$.

In this work, the initial spin parameter is set as $a_{*}=0.9$, and the starting time is set at the time when the initial BH mass (core mass) reaches 2.3 $M_{\odot}$.

\subsection{GWs from NDAFs}

Neutrino emission of NDAFs is mainly determined by the mass accretion rate and the properties of central BH \citep{Liu2016,Wei2019}. In the relativistic global solutions of NDAFs  \citep{Xue2013}, the detailed neutrino physics, the nuclear statistical equilibrium, and the conditions of the ignition and neutrino trapping are all considered. For the case of the viscosity parameter $\alpha = 0.1$, we can derive the fitting formula for the neutrino luminosity as a function of the BH mass and spin and the mass accretion rate, which is expressed as
\begin{align}
\log L_{\nu}\;(\rm{erg}\;s^{-1}) =\ &52.80-0.03m_{\rm{BH}}+1.01a_{*}
\notag
\\&+1.08\log \dot{m},
\end{align}
where $m_{\rm{BH}}=M_{\rm{BH}}/M_{\odot}$ and $\dot{m}=\dot{M}/M_{\odot}~\rm s^{-1}$ are the dimensionless BH mass and accretion rate, respectively. It should be noticed that the ignition and trapping accretion rates are roughly $\sim 0.001$ and $\sim 5 ~M_\odot ~\rm s^{-1}$, respectively, which also be significantly affected by the BH mass and spin and disk viscosity parameter \citep[e.g.,][]{Chen2007,Zalamea2011,Xue2013,Liu2017a}. But $\sim 0.01~M_\odot~\rm s^{-1}$ can be considered as the minimum accretion rate for detectable neutrino radiation from BH-NDAF systems in the center of massive collapsars \citep[e.g.,][]{Liu2016}.

\begin{figure}
\centering
\includegraphics[angle=0,scale=0.3]{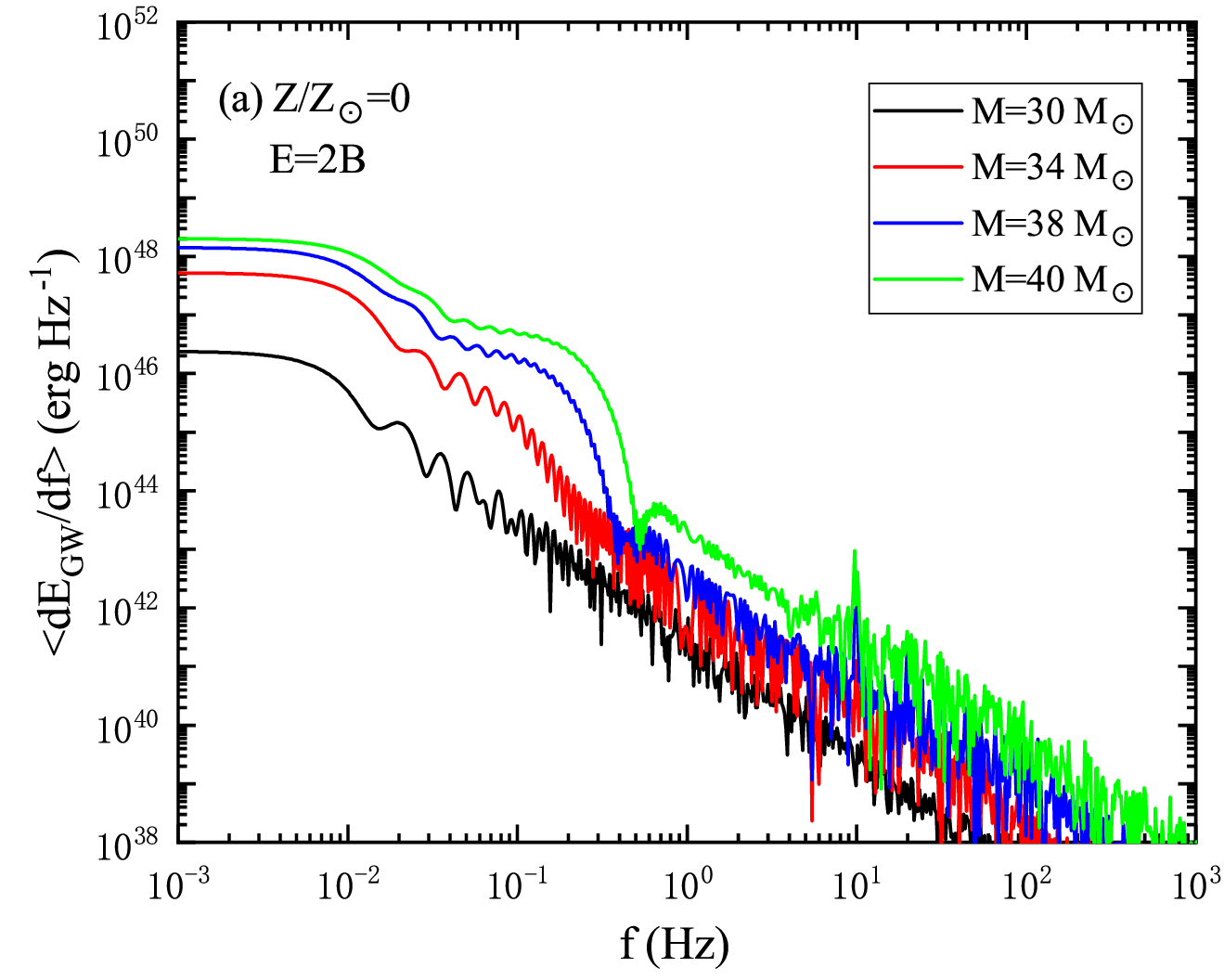}
\includegraphics[angle=0,scale=0.3]{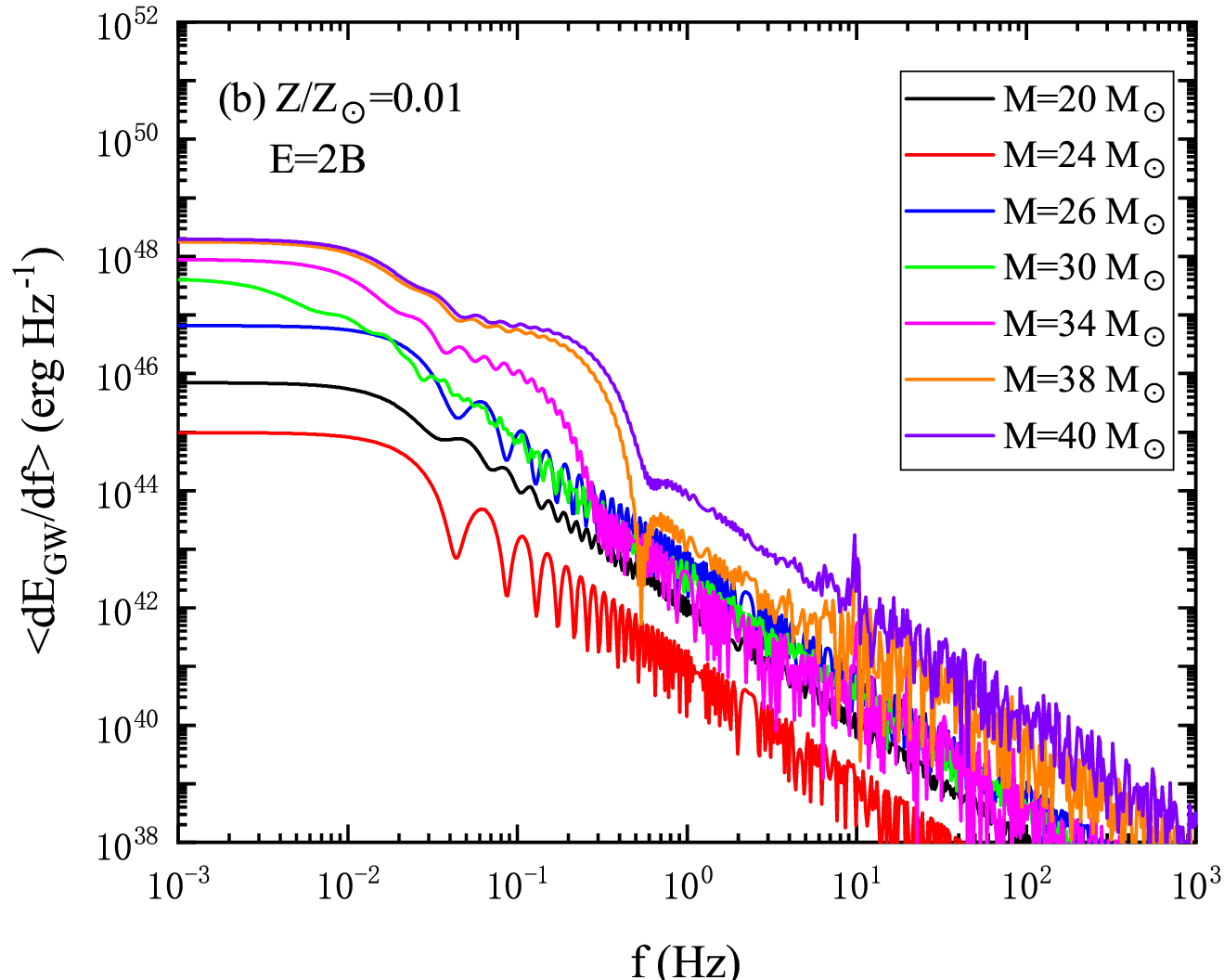}
\includegraphics[angle=0,scale=0.3]{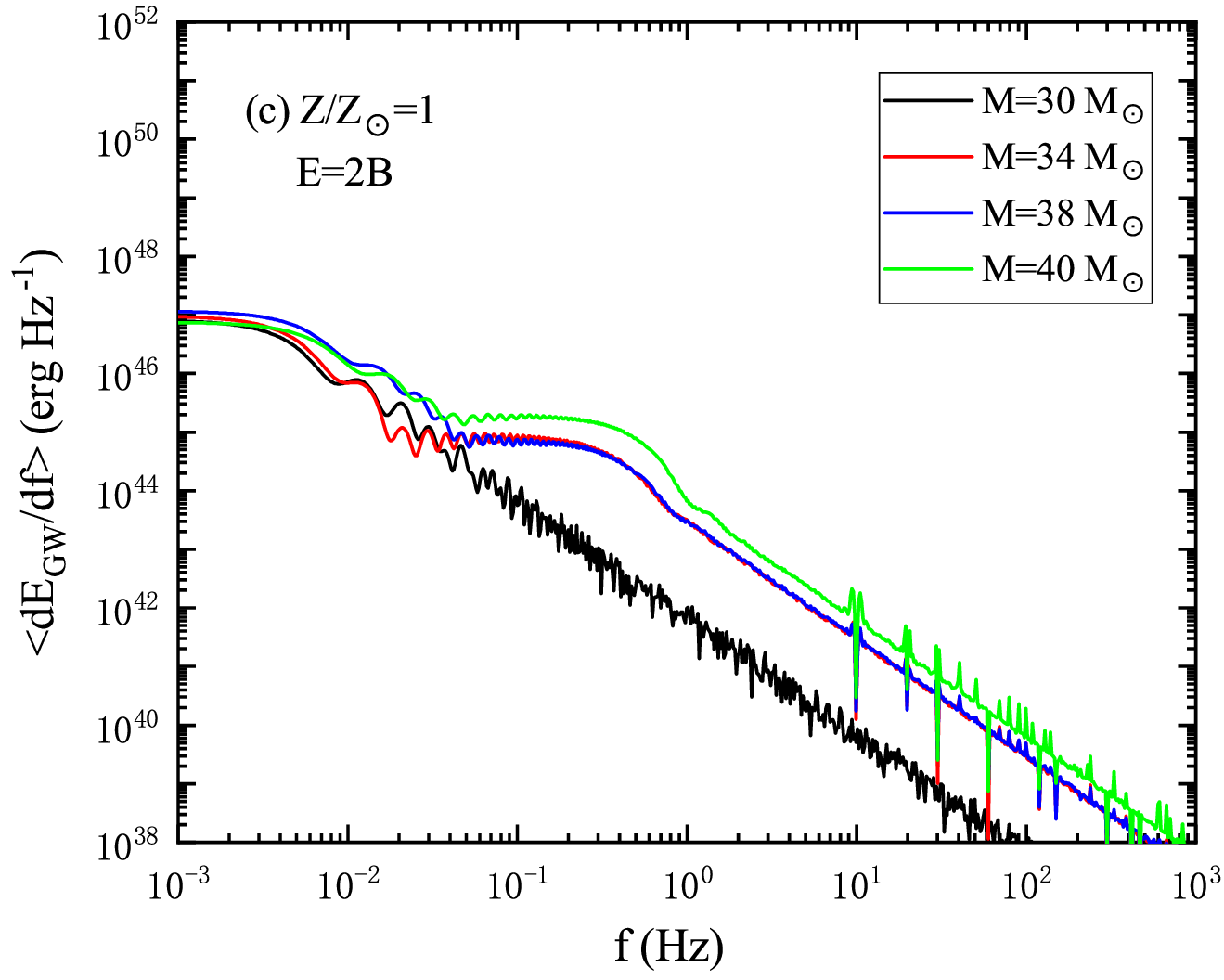}
\caption{The GW energy spectra of NDAFs with different progenitor masses and metallicites. Panels (a), (b), and (c) correspond to the cases of progenitors with metallicity of $Z/Z_{\odot}=0$, $0.01$, and $1$, respectively. The initial explosion energy is 2 B.}
\end{figure}

\begin{figure}
\centering
\includegraphics[angle=0,scale=0.3]{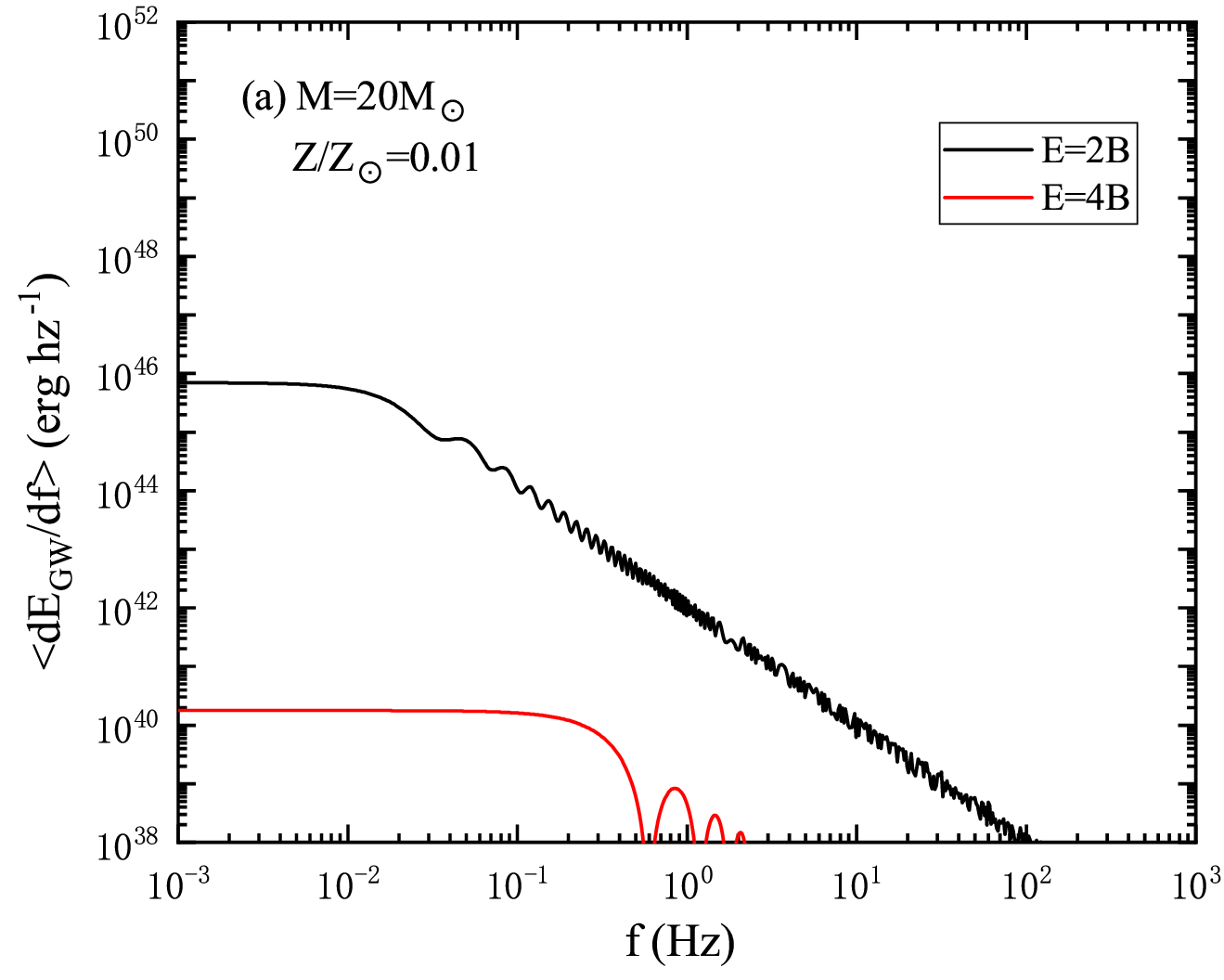}
\includegraphics[angle=0,scale=0.3]{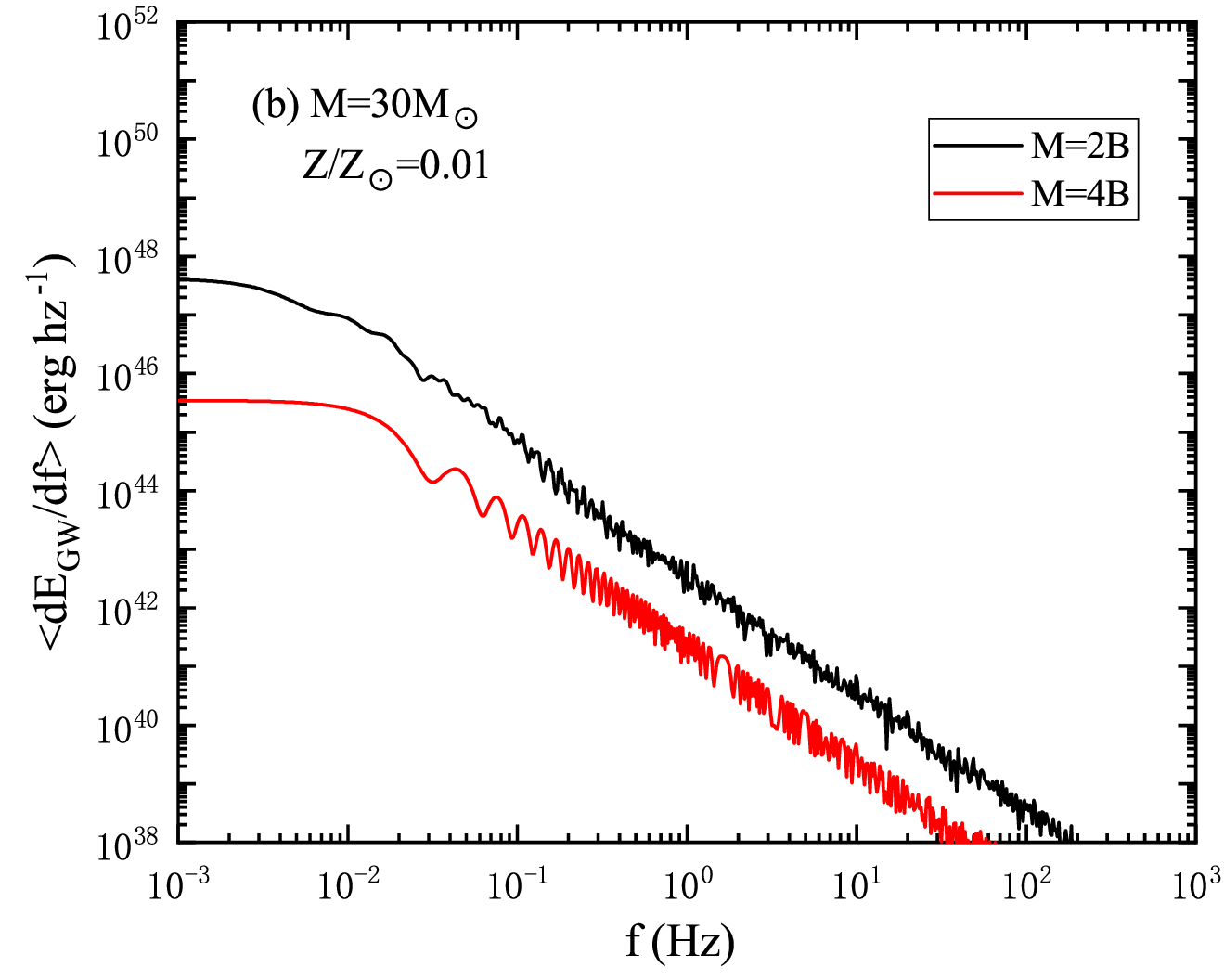}
\includegraphics[angle=0,scale=0.3]{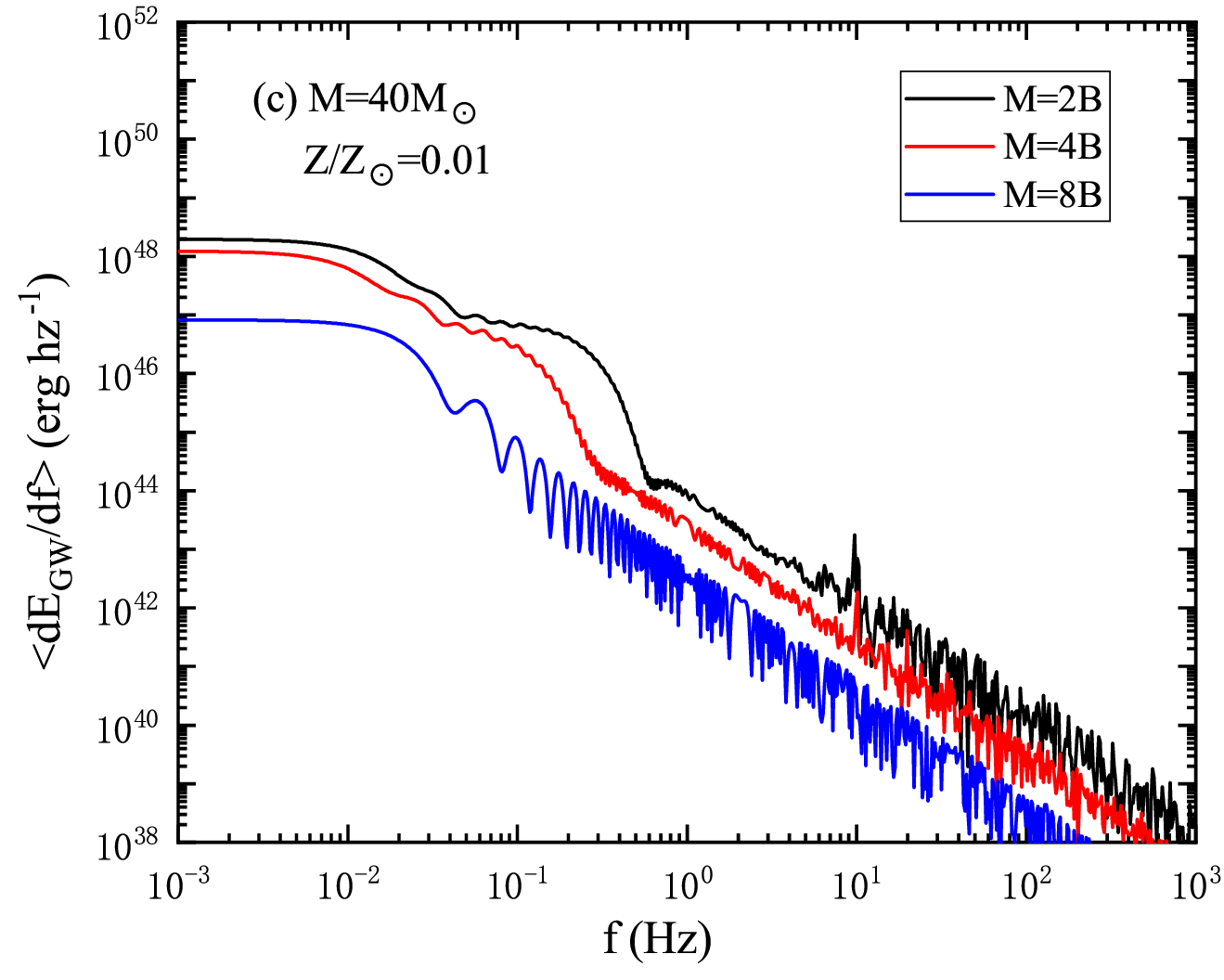}
\caption{The GW energy spectra of NDAFs with different initial explosion energies. Panels (a), (b), and (c) correspond to the cases of progenitors with mass of $20$, $30$, and $40$ $M_{\odot}$, respectively. The metallicity is set as $Z/Z_{\odot}=0.01$.}
\end{figure}

Here, we estimate the GWs from anisotropic neutrino emission, which is firstly proposed by \citet{Epstein1978}. We follow the formalism developed by
previous works \citep[e.g.,][]{Burrows1996,Mueller1997,Kotake2007,Suwa2009} to calculate the neutrino-induced GW signals. For a geometrically thin disk model for NDAFs, the local energy flux of GWs can be written as (for details, see \citet{Suwa2009})
\beq
\begin{split}
\frac{dE_{\rm{GW}}}{dA}&=\frac{G}{36\pi c^{5}D^{2}}(1+2cos\theta)^{2}tan^{4}(\frac{\theta }{2})\\
                         & \times\int_{-\infty }^{\infty }dt L_{\rm{\nu}}(t)^{2},
\end{split}
\eeq
where $dA=D^{2}d\Omega $ is the surface element, $D$ is the distance between the observer and the source, $\theta$ is the viewing angle. Here, $\theta=\pi/2$ corresponds to the case that the observer is located in the equatorial plane. Then, integrating over a sphere surrounding the source, one can calculate the total energy $E_{\rm{GW}}$. For $\theta=\pi/2$, the $E_{\rm{GW}}$ is calculated as
\beq
E_{\rm{GW}}=\frac{\beta G}{9c^{5}}\int_{-\infty }^{\infty }dt L_{\rm{\nu}}(t)^{2},
\eeq
where $\beta \sim 0.47039$. In order to calculate GW spectrum, one can write $L_{\rm{\nu}}(t)$ in terms of the inverse Fourier transform as
\beq
L_{\rm{\nu} }(t)=\int_{-\infty }^{+\infty} \tilde{L}_{\rm{\nu} }(f)e^{-2\pi ift}df;
\eeq
then, the GW energy spectrum can be deduced as
\beq
\frac{dE_{\rm{GW}}(f)}{df}=\frac{2\beta G}{9c^{5}}\left |  \tilde{L}_{\rm{\nu} }(f)\right |^{2}.
\eeq
In this work, we assume the orientation of disks is random and calculate the angle average GW energy spectrum per NDAF as
\beq
<\frac{dE_{\rm{GW}}(f)}{df}>=\sum_{i}\frac{\int_{\bigtriangleup \theta _{i}} d\theta}{\int_{0}^{\pi/2} d\theta } \frac{dE_{\rm{GW,i}}(f)}{df}
\eeq
where $\bigtriangleup\theta_{i}$ is the angle range of angle bin $i$, and $\frac{dE_{\rm{GW,i}}(f)}{df}$ is the observed GW energy spectrum at the corresponding viewing angle. In the subsequent calculations, we all adopt angle-averaged GW energy spectra.

\begin{table}
	\caption{$M_{\rm{min}}$ for different metallicities and initial explosion energies.}
	\label{table1}
	\begin{tabular}{ccc}
		\hline
		Metallicity   &  Initial Explosion Energy & $M_{\rm{min}}$    \\
        ($Z/Z_\odot$)   & ($\rm B$)     & ($M_\odot$)                 \\
        \hline
		 0	&	2	&	30	\\
         0.01	&	2	&	20	\\
         1	&	2	&	30	\\
         0.01	&	4	&	20	\\
         0.01	&	8	&	40	\\
		\hline
	\end{tabular}
\end{table}

The effects of the progenitor mass and metallicity on the GW energy spectrum of NDAFs are displayed in Figure 1. Panels (a), (b), and (c) correspond to the cases of progenitors with metallicity of $Z/Z_{\odot}=0$, $0.01$, and $1$, respectively. The initial explosion energy is 2 $\rm B$. In Panels (a) and (c), we only show GW energy spectra of NDAFs from progenitors with a mass greater than 30 $M_{\odot}$. In our simulations, for progenitors with metallicities of $Z/Z_{\odot}=0$ and $1$, the final remnant of progenitors with a mass less than $30M_{\odot}$ are NSs rather than BHs. The GW amplitude is mainly determined by the neutrino luminosity, which is related to the mass accretion rate. Most of the fallback and neutrino emission come from the core of the star. Therefore, the GW emission of the NDAF is determined by the compactness of the pre-SN star core. Some previous works \citep[e.g.,][]{OConnor2011,Sukhbold2014} studied the dependences of progenitor masses and metallicities on the structural characteristics of pre-SN stars and found a non-monotonic behavior for compactness as a function of progenitor mass and metallicity. As a result, for the same initial explosion energy, the GW emission of NDAFs is not strictly dependent on progenitor mass and metallicity. Besides, previous stellar evolutionary studies \citep{Sukhbold2014} found that the core compactness parameters of solar metallicity stars are commonly smaller than those of low metallicity stars. Thus, solar metallicity is unfavorable for neutrino emission and GW emission of NDAFs. Moreover, although the accretion rate is still large for massive progenitors, the central BH mass is more massive which will influence the neutrino luminosity of NDAFs \citep{Liu2021b}.

Figure 2 shows the effects of the initial explosion energy on the GW energy spectra of NDAFs. The metallicity is set to $Z/Z_{\odot}=0.01$. The black, red, and blue curves correspond to the initial explosion energies of 2, 4, and 8 $\rm B$, respectively. For different progenitor stars, the amplitudes of the spectral lines increase with decreasing initial explosion energy because the weaker explosion energy corresponds to a more powerful fallback and neutrino emission. In Figures 2(a) and 2(b), the case of 8 $\rm B$ is not shown because the final remnants of these progenitors are both NSs.

\section{SGWB from NDAFs}

A prediction of the SGWB requires a good understanding of the average GW energy spectrum and the event rate of NDAFs. The properties of progenitors and the initial explosion energy would determine whether the fallback process produces NDAFs and GW emission of NDAFs. Here, we investigate the effects of metallicities and initial explosion energies of progenitors on SGWB from NDAFs.

\subsection{Cosmic NDAF history}

\begin{figure}
\centering
\includegraphics[angle=0,scale=0.35]{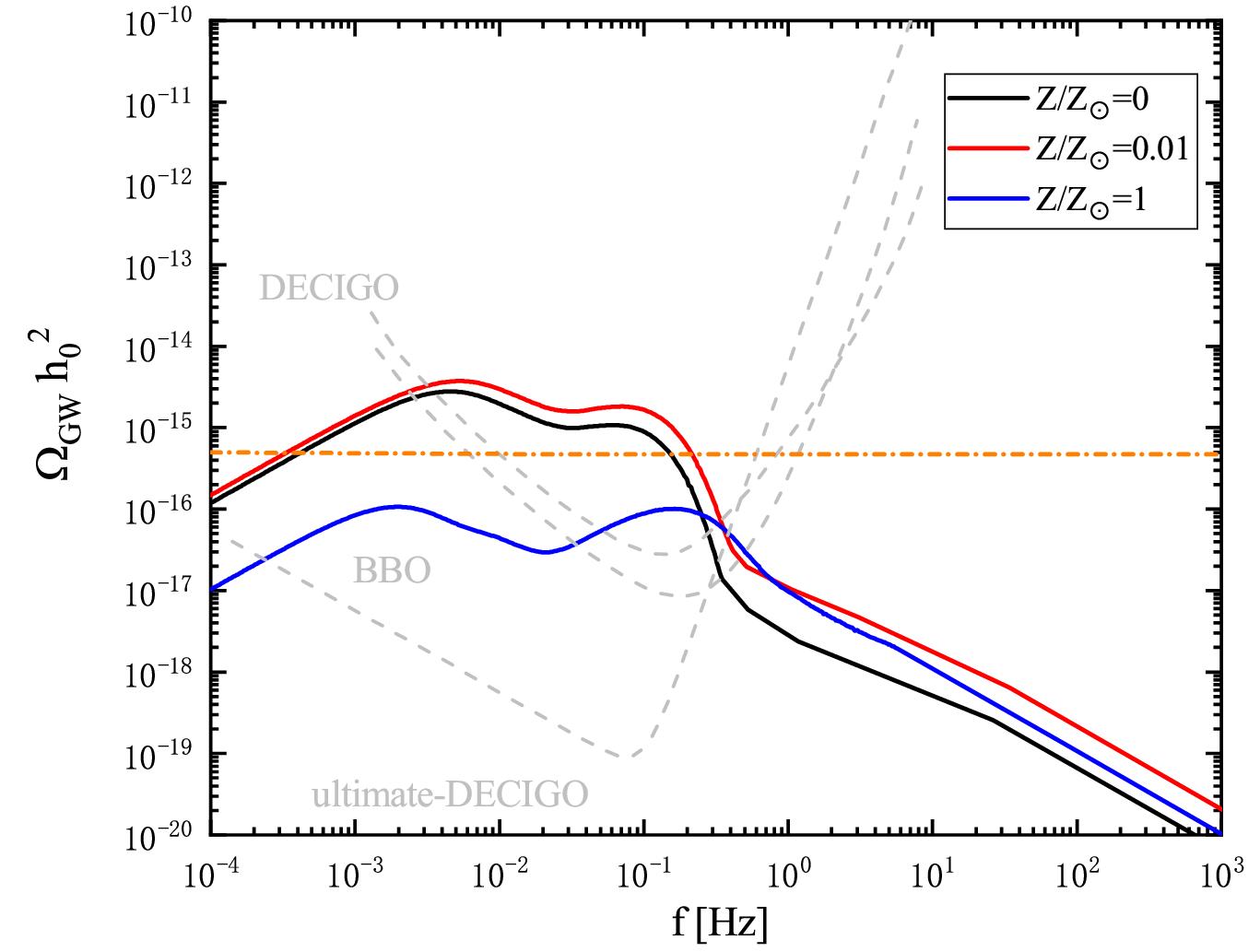}
\caption{The energy density parameter of SGWB for different metallicities with the initial explosion energy $E=2~\rm B$. The horizontal orange dashed line shows the SGWB generated during slow-roll inflation assuming $T/S=0.3$ for the ratio of the tensorial and scalar contributions to the cosmic microwave background radiation anisotropy and no running of the tensorial power-law index \citep{Turner1997}. The gray dashed lines represent the sensitivity curves of the indicated detectors.}
\end{figure}

The progenitor stars of CCSNe have relatively short lifetimes ($\lesssim 10^{8}$ years) compared to cosmic timescales \citep{Kennicutt1998}. Thus, the rate of NDAFs can be calculated using the star formation rate and IMF. The cosmic NDAF rate at a redshift of $z$ is calculated as
\beq
R_{\rm{NDAF}} (z)=R_{\rm{SFR}}(z)\frac{\int_{M_{\rm{min}}}^{M_{\rm{max}}} \Psi (M)dM}{\int_{0.1}^{125} M\Psi (M)dM},
\eeq
where $R_{\rm{SFR}}(z)$ is the cosmic star formation rate in units of $\rm{Mpc}^{-3}\; yr^{-1}$, which can be deduced from observations \citep[e.g.,][]{Hopkins2006,Reddy2008,Rujopakarn2010}. Here, we adopt the continuous broken power law description by \citet{Yuksel2008},
\beq
R_{\rm{SFR}} (z)= \dot{\rho} _{0}\left [ (1+z)^{\alpha\eta } +(\frac{1+z}{C} )^{\beta \eta } + (\frac{1+z}{D} )^{\gamma \eta }\right ]^{1/\eta },
\eeq
where $\alpha =3.4$, $\beta =-0.3$, $\gamma =-2$, $\eta =-10$, $C\simeq 5100$, $D\simeq 14$, and $\dot{\rho} _{0}=0.014 ~ \rm{Mpc}^{-3}~ yr^{-1}$. Here, we adopt a Salpeter IMF \citep{Salpeter1955} with $\Psi (M)\propto M^{\zeta}$ with $\zeta  =-2.35$ in the mass range of $0.1-125~M_{\odot}$, but explore a liberal range from $-2.15$ to $-2.45$ \citep{Bastian2010} in our final calculations. $M_{\rm{max}}$ and $M_{\rm{min}}$ are the maximum and minimum masses of progenitors that produce BH-NDAF systems, respectively. Due to the influence of the metallicity of the progenitor and initial explosion energy on the fallback process of CCSNe, not all massive stars can produce NDAFs. Especially, $M_{\rm{min}}$ depends on the metallicity of progenitor and initial explosion energy \citep{Liu2021}. According to the results of our simulations, we gave $M_{\rm{min}}$ for different metallicity and initial explosion energy, and the results are listed in Table \ref{table1}. Of course, we also checked that the fallback accretion rates of these progenitors are suitable for NDAFs until they decrease lower than the ignition accretion rates. The SGWB prediction depends weakly on the $M_{\rm{max}}$ and we set $M_{\rm{max}}\sim 50\;M_{\odot}$, which is the approximately upper limit of the progenitor mass for NDAFs because the new-born BH mass is too larger to effectively ignite NDAFs \citep{Liu2021b}.

\subsection{IMF-weighted average GW energy spectra}

\begin{figure}
\centering
\includegraphics[angle=0,scale=0.35]{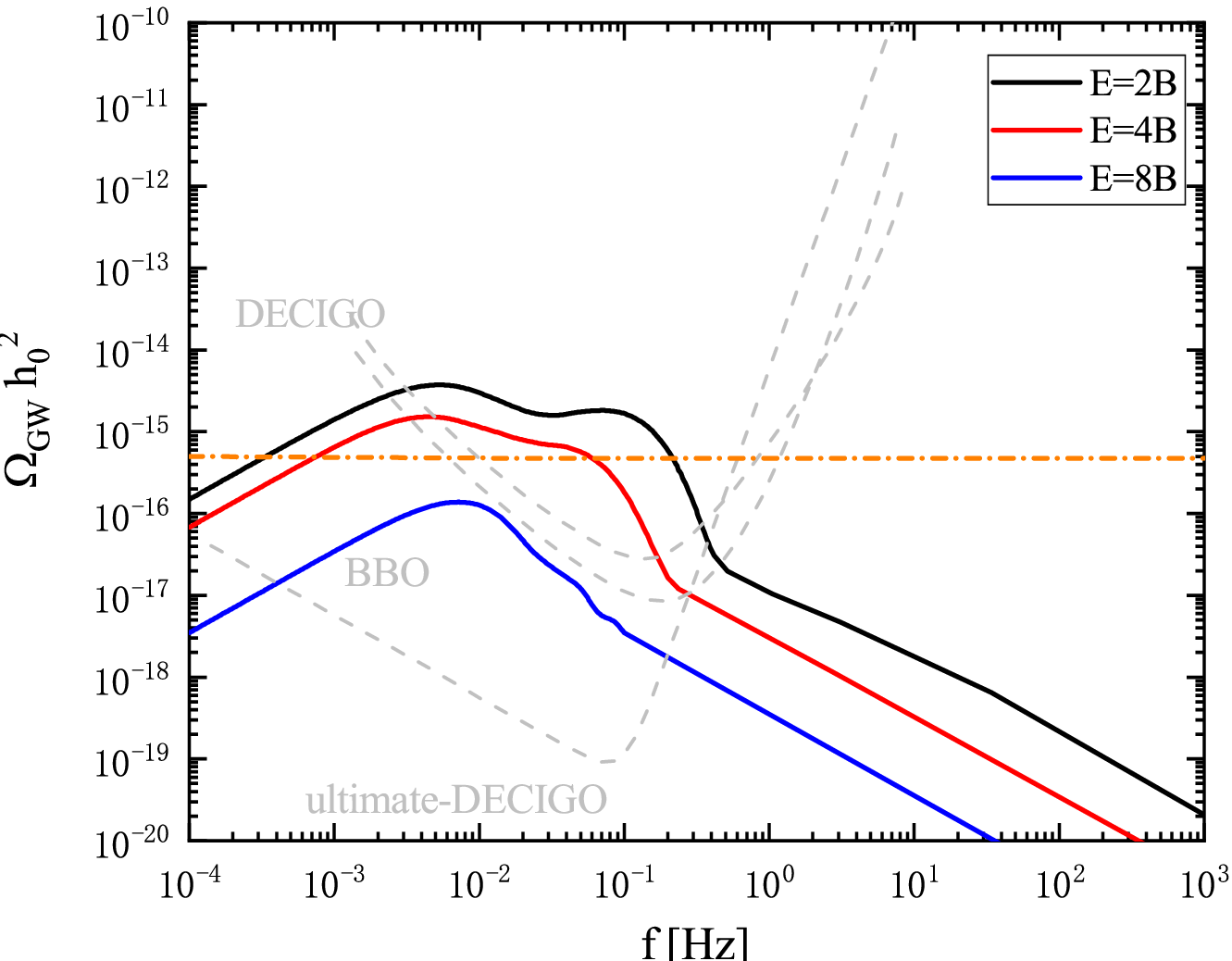}
\caption{The energy density parameter of SGWB for different initial explosion energies with metallicity $Z/Z_{\odot}=0.01$.}
\end{figure}

The average GW energy spectrum for a population of progenitors is computed by weighting each progenitor by
\beq
\frac{dE(f)}{df} =\sum_{i}\frac{\int_{\bigtriangleup M _{i}} \Psi (M)dM}{\int_{M_{\rm{min}}}^{M_{\rm{max}}} \Psi (M)dM} <\frac{dE_{{\rm{GW}}}(f)}{df}>_{i},
\eeq
where $\Psi (M)$ is once again the IMF, $\bigtriangleup M _{i}$ is the mass range of mass bin $i$, and $<\frac{dE_{{\rm{GW}}}(f)}{df}>_{i}$ is the angle-averaged GW energy spectrum of the NDAF with progenitor mass of $M_{i}$.

\subsection{Estimations of SGWB}

It is usually to characterize the SGWB by the energy density parameter \citep{Allen1999}, i.e.,
\beq
\Omega_{\rm{GW}}(f)=\frac{1}{\rho_{\rm{c}} }\frac{d\rho _{\rm{GW}}}{d\,{\rm{ln}} f},
\eeq
where $\rho_{\rm{GW}}$ is the gravitational energy density and $\rho_{\rm{c}}={3H_{0}^{2}}/{8\pi G}$ is the critical energy density needed to close
the universe. Then, following \citet{Phinney2001}, the sum of the energy densities radiated by a large number of independent NDAFs at each redshift is given as
\beq
\begin{split}
\Omega_{\rm{GW}}(f)&=\frac{1}{\rho_{\rm{c}}c^{2}}\\
                       & \times\int_{0}^{\infty}dz\frac{R_{\rm{NDAF}}(z)}{1+z}\left| \frac{dt}{dz}\right|f(z)\frac{dE(f_{z})}{df},
\end{split}
\eeq
where $f_{\rm{z}}\equiv f(1+z)$ and $\left |dt/dz \right |=[H_{0}(1+z)\sqrt{\Omega_{\Lambda}+\Omega _{m}(1+z)^{3}}]^{-1}$. Here, we ignore the anisotropy of GWs from NDAFs and adopt the cosmological parameters  $\Omega _{m}=0.3$, $\Omega_{\Lambda}=0.7$, and $H_{0}= 100 h_{0}\;\rm{km} \;\rm{s}^{-1}\;\rm{Mpc}^{-1}$ with $h_{0}=0.72$.

Figure 3 shows the effect of the metallicity of the progenitor on the spectrum of SGWB from NDAFs. The initial explosion energy of all progenitors is set as 2 B. The black, red, and blue curves correspond to $Z/Z_{\odot}=0$, $0.01$, and $1$, respectively. The gray dashed lines represent the sensitivity curves of DECIGO \citep{Yagi2017,Isoyama2018,Kawamura2021}, BBO \citep{Crowder2005, Corbin2006}, and the ultimate-DECIGO \citep{Seto2001}. Solar metallicity is not beneficial for the detection of the SGWB from NDAFs. This is because the solar metallicity is unfavorable for neutrino emission and GW emission of NDAFs. Moreover, if the solar metallicity is universal, the event rate of NDAFs would decrease. The horizontal orange dashed line shows the SGWB generated in the inflationary epoch. In low-frequency region ($\sim10^{-3}-10^{-1}$ Hz), our estimate for the SGWB from NDAFs can be comparable to the most optimistic SGWB from slow-roll inflation.

In Figure 4, we display the effects of the initial explosion energy on the spectrum of SGWB. We assume all progenitors have the same metallicity, $Z/Z_{\odot}=0.01$. The black, red, and blue curves correspond to the initial explosion energy of 2, 4, and 8 B, respectively. The weaker explosion energy corresponds to a more powerful fallback, which enhances the GW emission and the event rate of NDAFs. As a result, the initial explosion energy will significantly affect the detection of the SGWB from NDAFs.

\begin{figure}
\centering
\includegraphics[angle=0,scale=0.35]{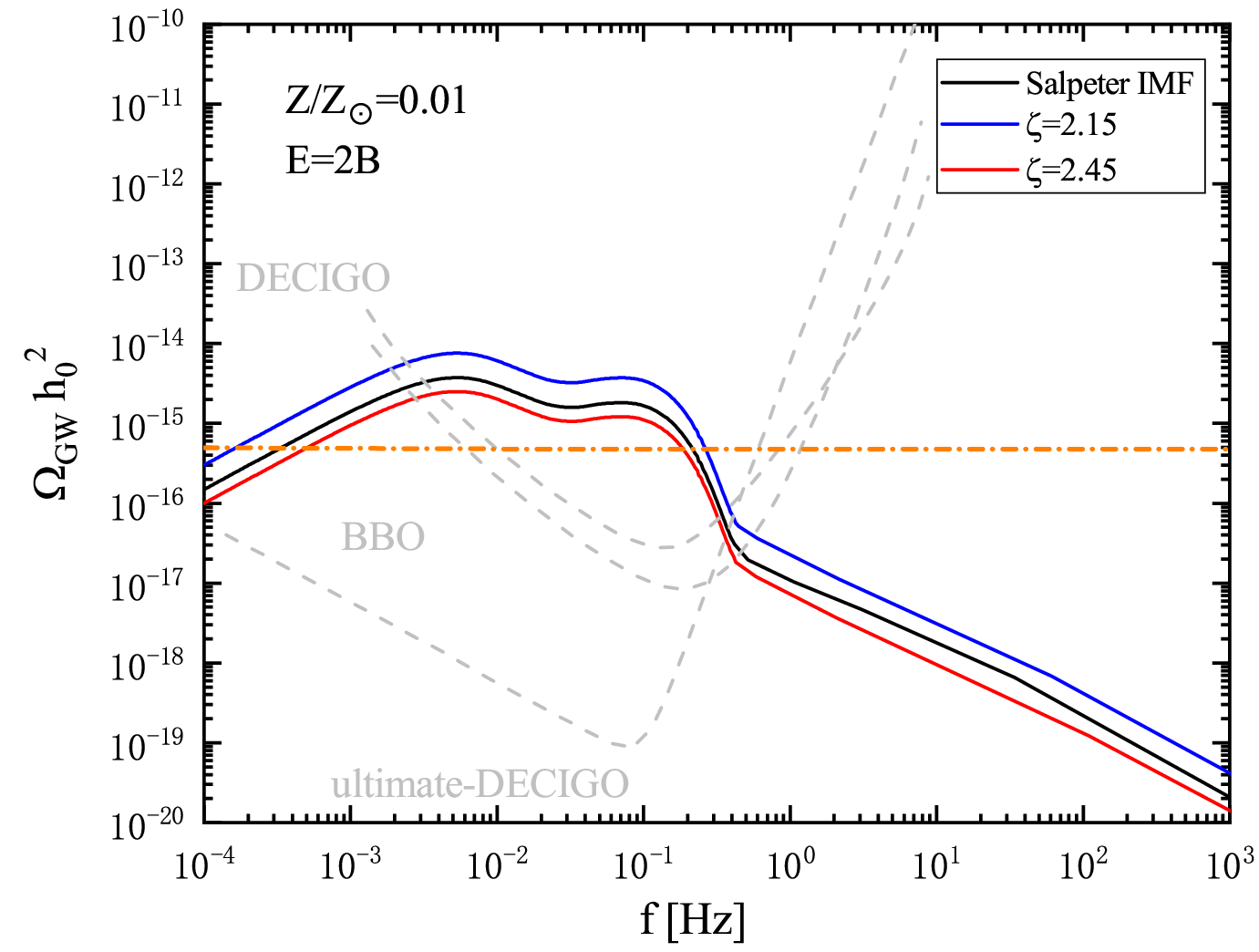}
\caption{The energy density parameter of SGWB for different IMF. The initial explosion energies is 2 B and metallicity is $Z/Z_{\odot}=0.01$.}
\end{figure}

In Figure 5, we explore the effect of IMF on the spectrum of SGWB from NDAFs. The black curve corresponds to the Salpeter IMF with $\zeta =-2.35$. The red and blue curves correspond to $\zeta =-2.15$ and $\zeta =-2.45$, respectively. A shallower IMF (i.e., $\zeta =-2.15$) is beneficial for the detection of SGWB from NDAFs. The SGWB differs by almost a factor of 3 between the case of $\zeta =-2.15$ and the case of $\zeta =-2.45$. As shown in Figure 1, for the same progentior metallicity and initial explosion energy, relatively higher mass stars correspond to the higher accretion rate, then the more powerful GW emission of NDAFs until the effects of the central BH mass dominate the ignition of NDAFs. Shallower IMFs have more high-mass stars which contribute to the SGWB.

\section{Conclusions and Discussion}

In this paper, we estimate the SGWB produced by NDAFs based on fallback CCSN simulations. The effects of progenitor properties and initial explosion energies on SGWB from NDAFs are studied. These factors can affect the GW energy spectra and event rates of NDAFs. The GW emission of NDAFs is determined by the fallback process in CCSNe. Lower initial explosion energy is beneficial for producing a more powerful fallback, resulting in stronger neutrino emission and GW emission of NDAFs. Moreover, low initial explosion energy may enhance the event rates of NDAFs. Therefore, if low initial explosion energy is universal CCSNe, even for failed CCSNe, the SGWB from NDAFs may be detected by interferometers such as DECIGO and BBO. The influence of metallicity on SGWB from NDAFs is not monotonic because the compactness of the pre-SN star core is not strictly dependent on metallicity. However, solar metallicity is not beneficial for the detection of SGWB. Another uncertainty in our models is the IMF. A shallower IMF would increase the amplitude of SGWB from NDAFs. Therefore, the SGWB could be a valuable tool to investigate the IMF.

In NDAF models, we adopt $\alpha=0.1$ as a typical viscosity parameter. Actually, a disk with low $\alpha$ is denser and has a higher neutrino luminosity \citep[e.g.,][]{Popham1999,Liu2007,Chen2007,Liu2017a}. Thus, the GW emission of NDAFs with low $\alpha$ might be more powerful. Besides, in our calculations, we do not consider the disk outflows, which plays an important role in critical accretion systems \citep[e.g.][]{Liu2008,Liu2014,Gu2015}. The neutrino luminosity of NDAFs with strong disk outflows would be at least one order of magnitude lower than that of NDAFs without outflows \citep{Liu2021b}. As a result, powerful disk outflow will decrease the GW emission of NDAFs. If powerful disk outflows are universal in NDAFs, the amplitude of SGWB from NDAFs would decrease.

In summary, the uncertainties of the SGWB from NDAFs are large. The detection of the corresponding neutrino background will lead to an improved prediction of the SGWB from NDAFs. In \citet{Wei2024}, we investigated the diffuse NDAF neutrino background (DNNB) in detail. Metallicity and initial explosion energy have similar effects on both DNNB and SGWB from NDAFs. For the optimistic cases where the typical initial explosion energy is low, the DNNB might be detected by the upcoming larger neutrino detectors such as Hyper-Kamiokande, Jiangmen Underground Neutrino Observatory (JUNO), and Deep Underground Neutrino Experiment (DUNE). The detection of DNNB would help constrain the average neutrino emission and event rate of NDAFs. By combining the DNNB and SGWB, we would better understand CCSNe and NDAFs.

It is noted that the SGWB from NDAFs might mask the SGWB generated in the inflationary epoch at low frequency. Besides, the SGWB from cosmological CCSNe and Pop III stars are expected to mask the inflationary GWs in some range of frequencies. These astrophysical foreground sources could be a significant problem for searches of the inflationary SGWB. It is necessary to disentangle these foreground sources from the inflationary SGWB in future detection.

\acknowledgments
We thank Prof. Alexander Heger for providing us pre-SN data. We also thank Xing-Jiang Zhu for helpful discussion. This work was supported by the National Key R\&D Program of China (Grant No. 2023YFA1607902), the National Natural Science Foundation of China (Grant Nos. 12173031, 12221003, and 12303049), the Postdoctoral Fellowship Program of China Postdoctoral Science Foundation (Grant No. GZC20231424), and the science research grants from the China Manned Space Project.

\end{document}